\documentclass[reprint,amsmath,amssymb,aps,pra,superscriptaddress,nofootinbib,floatfix
]{revtex4-2}

\usepackage{graphicx}
\usepackage{dcolumn}
\usepackage{bm}
\usepackage{hyperref}
\usepackage{amsmath}

\usepackage[
    protrusion=true,
    activate={true,nocompatibility},
    final,
    tracking=true,
    kerning=true,
    spacing=true,
    spacing=nonfrench,
    factor=1000]{microtype}
\hypersetup{
 colorlinks=true, 
 linkcolor=blue, 
 citecolor=blue,        
 filecolor=blue,      
 urlcolor=blue}
 \usepackage{xcolor}
 \usepackage{ulem}

\begin{document}

\preprint{APS/123-QED}


\title{Freeform gradient-index media reconstruction}

\author{V. Ch\'{a}vez-Islas}
\affiliation{Instituto Nacional de Astrof\'{i}sica, \'{O}ptica y Electr\'{o}nica, Coordinaci\'{o}n de \'{O}ptica, Tonantzintla, Puebla 72840, Mexico}
\author{F. H. Maldonado-Villamizar}%
\affiliation{SECIHTI - Instituto Nacional de Astrof\'{i}sica, \'{O}ptica y Electr\'{o}nica, Coordinaci\'{o}n de \'{O}ptica, Tonantzintla, Puebla 72840, Mexico}
\author{E. Esp\'{i}ndola-Ramos}%
\affiliation{Instituto Nacional de Astrof\'{i}sica, \'{O}ptica y Electr\'{o}nica, Coordinaci\'{o}n de \'{O}ptica, Tonantzintla, Puebla 72840, Mexico}
\author{S. Ch\'{a}vez-Cerda}%
\affiliation{Instituto Nacional de Astrof\'{i}sica, \'{O}ptica y Electr\'{o}nica, Coordinaci\'{o}n de \'{O}ptica, Tonantzintla, Puebla 72840, Mexico}
\author{J. E. G\'{o}mez-Correa}%
\email{jgomez@inaoep.mx}
\affiliation{Instituto Nacional de Astrof\'{i}sica, \'{O}ptica y Electr\'{o}nica, Coordinaci\'{o}n de \'{O}ptica, Tonantzintla, Puebla 72840, Mexico}


%
%
%

\begin{abstract}
Reconstructing a gradient-index (GRIN) medium consists of determining a spatial refractive index distribution that produces a prescribed ray trajectory, an inverse problem in geometrical optics. Previous approaches have used geometrical symmetries and conserved ray quantities to address this inverse problem. Extending reconstruction to freeform media requires approaches that do not rely on these symmetries. Here, we present a three-dimensional point-wise method for reconstructing freeform GRIN media through the inverse application of local refraction. The method does not require geometrical symmetry, paraxial approximations, or prior knowledge of the iso-indicial surfaces, and is straightforward to implement. Reconstruction using two interface geometries demonstrates that different refractive index distributions can be obtained from the same prescribed trajectory function. The choice of interfaces thus provides an additional degree of freedom for optical design, subject to local refraction conditions and compatibility between overlapping reconstructed regions.
\end{abstract}

\maketitle


\section{Introduction}

Gradient-index (GRIN) media are characterized by a spatially varying refractive index distribution that continuously bends light rays during propagation \cite{GomezLibro,Lakshminarayanan}. Their internal structure can be described in terms of iso-indicial surfaces \cite{Gomez-Correa:21}, i.e., imaginary three-dimensional layers within an inhomogeneous optical medium where every point shares the same refractive index. According to the geometry of these surfaces, GRIN media can be broadly classified as symmetric or freeform \cite{GomezCorrea2023}. In symmetric GRIN media, the iso-indicial surfaces exhibit a well-defined geometrical symmetry, as in the classical axial, cylindrical, and spherical distributions \cite{Lakshminarayanan}. Freeform GRIN media, in contrast, are characterized by iso-indicial surfaces without a predefined symmetry, allowing the refractive index to vary arbitrarily in three-dimensional space \cite{Lippman:21}.

In geometrical optics, determining the ray trajectory through a known GRIN distribution constitutes the direct problem and has been extensively addressed through analytical and numerical approaches \cite{Marchand:70, Marchand:88, Marchand:72, Moore:75, Streifer:71, Paxton:71, Buchdahl:73, Liu:23, Sharma:82, Sharma:85, Sakamoto:87, Gomez-Correa:21, Gomez-Correa:22, Chavez-Islas:25, Chavez2026generalized}. Conversely, the inverse problem consists of determining a refractive index distribution capable of producing a prescribed ray trajectory and remains an active area of research \cite{Nemoto:79,Minano:85,Borghero_2005,Borghero:06,Borghero:11,Borghero:16,Manipulating,GRINMapping,GRINMapping2,GomezCorrea2023}. 

Early approaches derived refractive index distributions producing prescribed two-dimensional ray paths and showed that different index distributions may support the same prescribed trajectory, revealing the nonunique nature of the inverse problem \cite{Nemoto:79}. 

More recently, a reconstruction method for symmetric GRIN media based on Fermat ray invariants was introduced \cite{GomezCorrea2023}. These conserved quantities arise from geometrical symmetries and can be interpreted as generalized forms of Snell's law \cite{Gomez-Correa:21,Gomez-Correa:22}, allowing the refractive index to be recovered along a prescribed trajectory without integrating the ray differential equation. This approach provides a physical connection to freeform reconstruction: although a corresponding symmetry-based invariant is generally unavailable, the local refraction law underlying the reconstruction remains applicable.



In this paper, we address the inverse problem for three-dimensional freeform GRIN media using a point-wise reconstruction approach. The proposed method requires neither geometrical symmetry nor paraxial approximations or prior knowledge of the iso-indicial surfaces. Its implementation relies on the successive application of a simple local refraction relation, without integrating the ray differential equation. We show how the nonuniqueness of the inverse problem manifests through the choice of reconstruction interfaces: different interface geometries can yield different refractive index distributions for the same prescribed trajectory. This geometrical flexibility provides an additional degree of freedom for tailoring the reconstructed distribution, subject to the local refraction conditions.
%


\section{Reconstruction method}

The reconstruction starts from a prescribed ray trajectory and a selected family of surfaces whose geometry is known. The reconstructed refractive index distribution depends on this choice, since the surfaces determine the intersection points and local normals along the prescribed trajectory. Their intersections divide the trajectory into consecutive segments, which are approximated as straight propagation paths connected by local refraction events. At each intersection, the directions of the adjacent segments and the surface normal provide the geometrical information needed to recover the refractive index ratio through Snell’s law. Starting from a known input refractive index, this relation is applied successively to construct a piecewise-homogeneous representation of the medium along the prescribed trajectory. 
\begin{figure}[t]
\centering
\includegraphics[width=\linewidth]{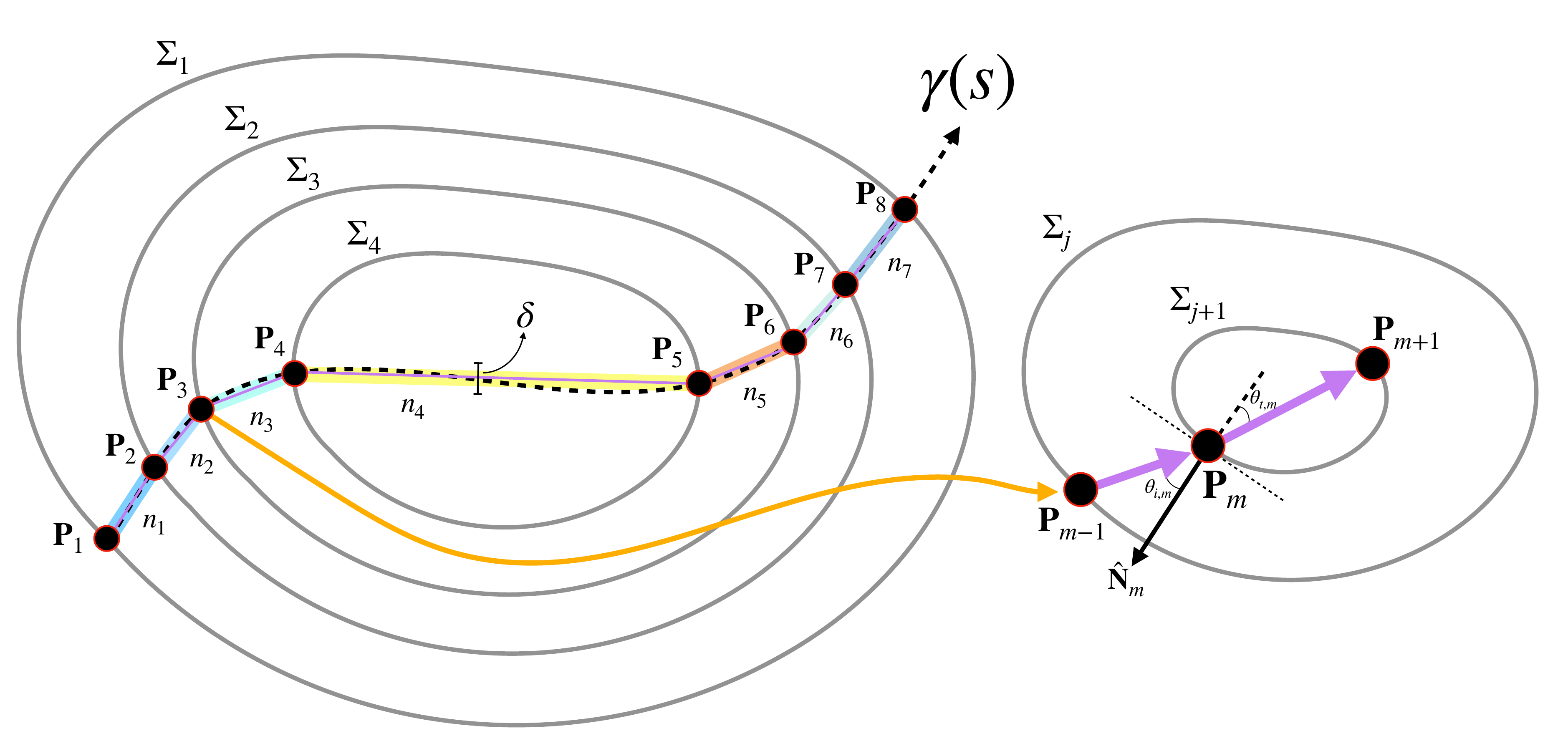}
\caption{Schematic representation of the reconstruction method. The prescribed trajectory $\boldsymbol{\gamma}(s)$ intersects the selected interfaces $\Sigma_j$ at points $\mathbf{P}_m$, ordered along the propagation direction. Consecutive intersections define locally homogeneous segments with refractive index $n_m$ and transverse extent controlled by $\delta$. The enlarged view shows the local normal $\hat{\mathbf{N}}_m$ and the incidence and transmission angles used to recover $n_m$ through Snell's law.}
    \label{fig:reconstruccion}
\end{figure}

Let $\boldsymbol\gamma(s)$ denote the prescribed ray trajectory, where $s$ denotes a trajectory parameter, and let $\{\Sigma_j\}_{j=1}^{J}$ be a selected family of reconstruction interfaces, where $J$ is the total number of interfaces, see Fig. \ref{fig:reconstruccion}. These interfaces specify the geometry at which local refraction conditions are imposed and are not constrained to be iso-indicial surfaces: the refractive index is recovered locally at each ray–interface intersection and may vary across a given interface. Each interface is described by the parametric surface
\begin{equation} \label{eq:s_param}
  \mathbf\Sigma_j(u,v) = \begin{bmatrix}
    X_j(u,v)\\
    Y_j(u,v)\\
    Z_j(u,v)
  \end{bmatrix},
\end{equation}
where $u$ and $v$ are the surface parameters, and $X_j(u,v)$, $Y_j(u,v)$, and $Z_j(u,v)$ are the Cartesian coordinates of the $j$-th interface.

The prescribed trajectory intersects the reconstruction interfaces at points
\begin{equation} 
\label{eq:intersecciones}
  \mathbf{P}_m=\boldsymbol{\gamma}(s_m)=\mathbf{\Sigma}_{j_m}(u_m,v_m),  
\end{equation}
where $\mathbf{P}_m$ is the $m$-th intersection point along the ray, $s_m$ is the corresponding trajectory parameter, and $j_m$ identifies the interface crossed at that point. The parameters $(u_m,v_m)$ locate $\mathbf{P}_m$ on that interface. The intersection points are ordered by increasing $s_m$, following the direction of propagation. Thus, $m$ labels successive intersections, whereas $j$ labels the reconstruction interfaces. This distinction allows the same interface to be crossed more than once, with each crossing assigned a different value of $m$.

At each intersection $\mathbf{P}_m$, the local unit normal is calculated from the interface parametrization as
\begin{equation}
\label{eq:normal}
\hat{\mathbf{N}}_m=
\frac{
\partial_u\mathbf{\Sigma}_{j_m}(u,v)
\times
\partial_v\mathbf{\Sigma}_{j_m}(u,v)
}{
\left\|
\partial_u\mathbf{\Sigma}_{j_m}(u,v)
\times
\partial_v\mathbf{\Sigma}_{j_m}(u,v)
\right\|
}\Bigg|_{u=u_m,\,v=v_m},
\end{equation}
where the notation $\partial_\alpha\mathbf{\Sigma}_{j_m}(u,v)$ denotes the partial derivative of $\mathbf{\Sigma}_{j_m}(u,v)$ with respect to $\alpha \in \{u,v\}$.
The adjacent trajectory segments define the incident and transmitted unit vectors,
\begin{equation}
\label{eq:directions}
\hat{\mathbf{i}}_m=
\frac{\mathbf{P}_m-\mathbf{P}_{m-1}}
{\left\|\mathbf{P}_m-\mathbf{P}_{m-1}\right\|},
\qquad
\hat{\mathbf{t}}_m=
\frac{\mathbf{P}_{m+1}-\mathbf{P}_m}
{\left\|\mathbf{P}_{m+1}-\mathbf{P}_m\right\|}.
\end{equation}
Together with $\hat{\mathbf{N}}_m$, these vectors determine the incidence and transmission angles through
\begin{equation}
\label{eq:angulos}
\sin \theta_{i,m} = \left\Vert{} \mathbf{\hat{i}}_m \times \mathbf{\hat{N}}_m \right\Vert{}, \quad \sin \theta_{t,m} = \left\Vert{} \mathbf{\hat{t}}_m \times \mathbf{\hat{N}}_m \right\Vert{}
\end{equation}

\begin{figure}[t]
\centering
\includegraphics[width=\linewidth]{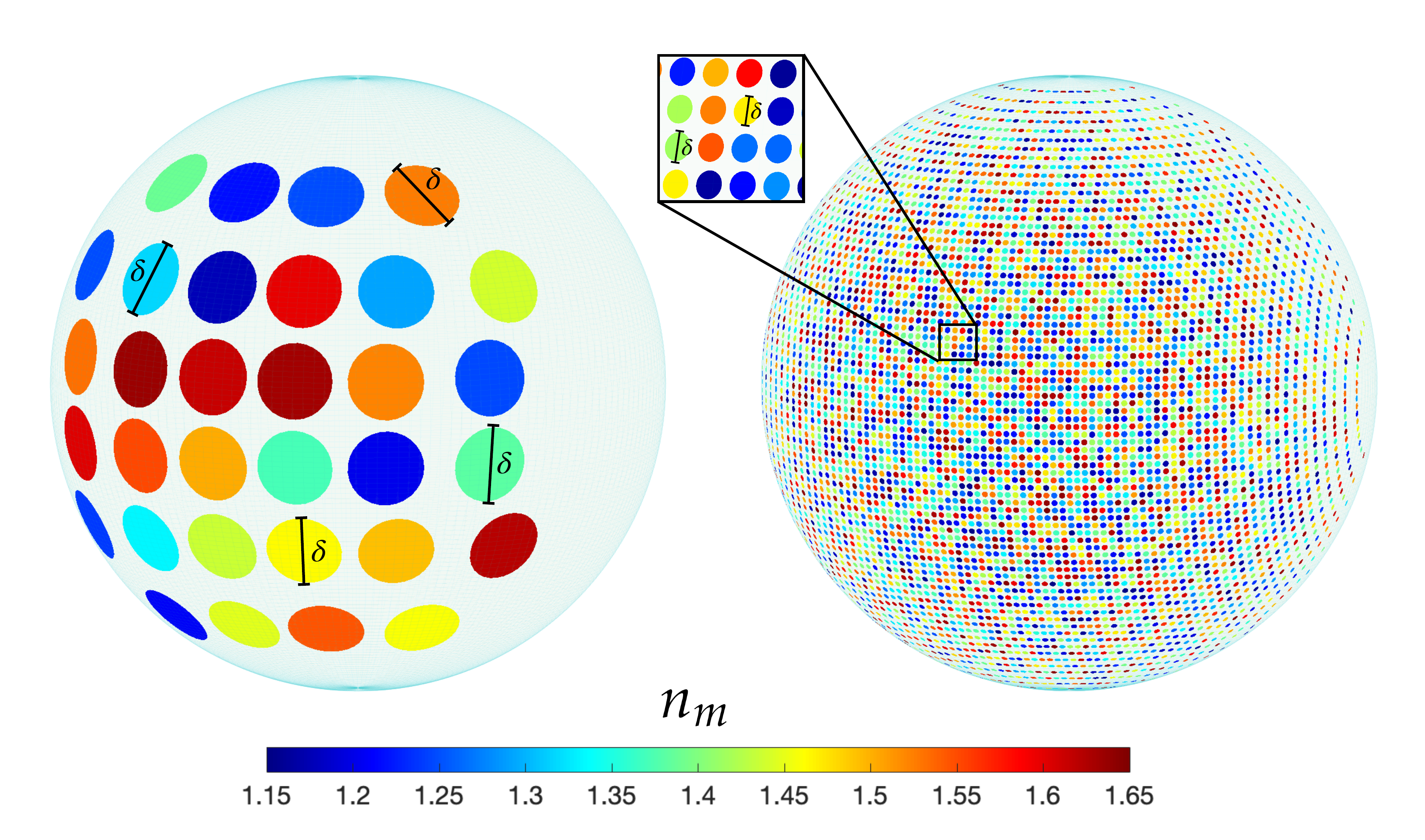}
\caption{Local refractive index assignment on a reconstruction interface for larger (left) and smaller (right) values of $\delta$. Each colored patch represents a region assigned a constant refractive index $n_m$, with its transverse extent controlled by $\delta$. Different regions on the same interface may have different refractive indices, as indicated by the color scale. The inset shows an enlarged view of the smaller patches.}
    \label{fig:Delta}
\end{figure}

For these directions to describe a transmitted ray, they must satisfy the local vectorial refraction conditions: the incident direction, transmitted direction, and interface normal must be coplanar, the tangential components of the two directions must have the same orientation, and their normal components must correspond to propagation across the interface. Once these conditions are satisfied, Snell's law gives the refractive index $n_m$ required at $\mathbf{P}_m$:
\begin{equation}
\label{eq:reconstruction}
n_m=n_{m-1}\frac{\sin\theta_{i,m}}{\sin\theta_{t,m}},
\end{equation}
where $n_{m-1}$ is the refractive index immediately before the intersection, and $n_m$ corresponds to the segment extending from $\mathbf{P}_m$ to $\mathbf{P}_{m+1}$. Disks of diameter $\delta$ on consecutive interfaces define a locally homogeneous pillbox-shaped region traversed by this ray segment. Each disk is defined as a circle in the local tangent plane of the interface, independently of the surface geometry and the ray’s incidence angle. The parameter $\delta$ can be adjusted to control the transverse extent of the local refractive index assignment. The interfaces delimit the longitudinal extent of the region, and all points within this local volume are assigned the same refractive index value $n_m$. This pillbox-based assignment avoids imposing a uniform index over an entire layer and allows different regions of the optical component to be tailored to distinct prescribed trajectories, provided that the assigned refractive index values are equal wherever the local volumes overlap. Figure \ref{fig:Delta} shows the reconstructed regions for larger and smaller values of $\delta$.

\section{Reconstruction accuracy}

To assess the reconstruction accuracy, we consider a Luneburg lens \cite{Luneburg}, for which both the refractive index distribution and the ray trajectories are analytically known \cite{Luneburg,Grzesik}. Although this lens is spherically symmetric, it provides a well-defined benchmark: the analytical trajectories serve as inputs to the reconstruction method, and the recovered refractive indices can be compared directly with the exact distribution. Its refractive index distribution is given by \cite{Luneburg}
\begin{equation}
\label{eq:luneburg}
n(r)=\sqrt{2-r^2},
\end{equation}
where $r\in[0,1]$ is the radial coordinate normalized by the lens radius, with the origin at the center of the lens. The ray trajectories inside the lens are described analytically by \cite{Grzesik}
\begin{equation}
y(x)=y_1\cos\!\left[
\sin^{-1}\!\left(\frac{x}{\Lambda_1}\right)
-\sin^{-1}\!\left(\frac{x_1}{\Lambda_1}\right)
\right],
\label{eq:luneburg_trajectory}
\end{equation}
where $\Lambda_1=\sqrt{1+x_1^2}$, and $(x_1,y_1)$
is the ray entry point in its propagation plane. 
\begin{figure}[ht]
\centering
\includegraphics[width=\linewidth]{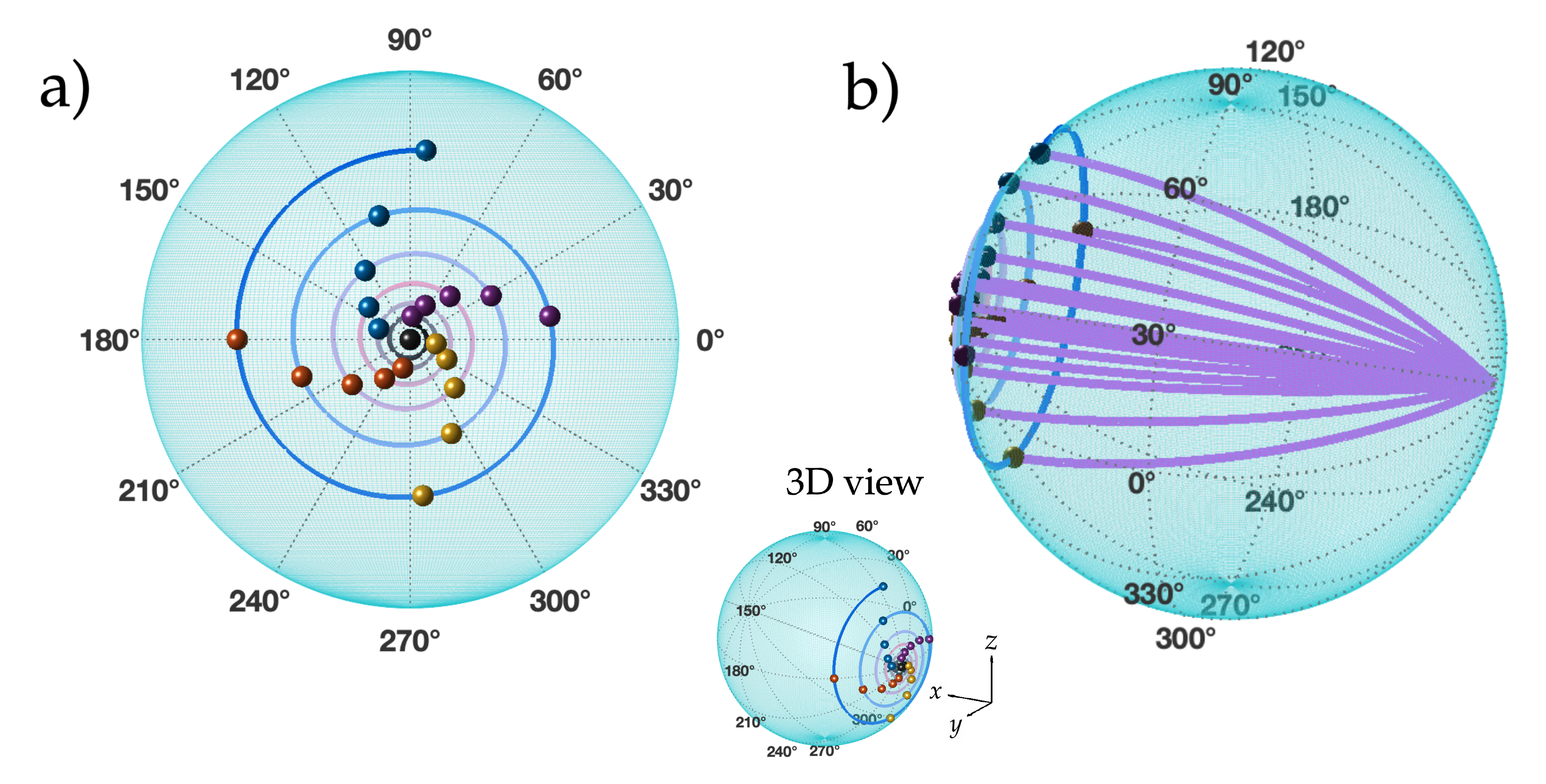}
\caption{Prescribed rays used for the Luneburg reconstruction. (a) Anterior hemisphere showing the entry points distributed along an exponential spiral, sampling progressively smaller distances from the optical axis at different azimuths; the inset shows their three-dimensional arrangement. (b) Corresponding analytical ray trajectories inside the lens.}
    \label{fig:Spiral}
\end{figure}

Before quantifying the reconstruction accuracy, we first demonstrate the method using 20 prescribed ray trajectories whose entry points are distributed along an exponential spiral on the surface of the anterior hemisphere, as shown in Fig.~\ref{fig:Spiral}(a). This arrangement varies both the radial distance from the optical axis and the azimuth, placing successive entry points progressively closer to the axis while avoiding repeated radial positions \cite{Navarro}. Thus, the selected rays sample entry points at different distances from the lens axis and different azimuthal angles using a limited number of trajectories. The corresponding analytical trajectories, each lying in its respective meridional plane, are shown in Fig.~\ref{fig:Spiral}(b).

\begin{figure}[t]
\centering
\includegraphics[width=\linewidth]{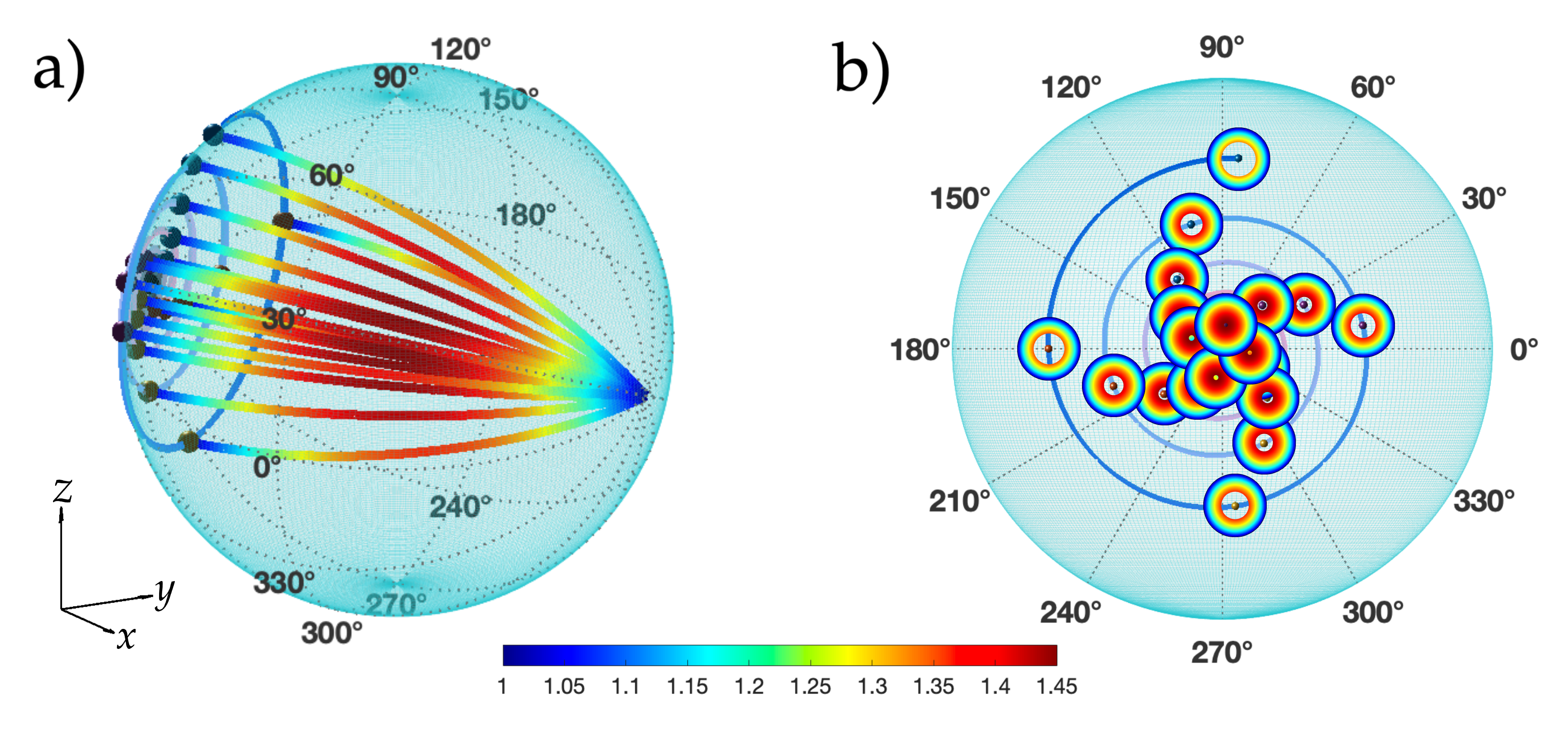}
\caption{Reconstruction of the Luneburg refractive index distribution using $J=300$ concentric spherical interfaces. (a) Recovered refractive index along the prescribed trajectories. (b) Radial profiles recovered from individual rays, extended using the known spherical symmetry and displayed at their corresponding entry positions. Rays passing closer to the lens center recover a wider radial interval. Colors represent the reconstructed refractive index values.}
    \label{fig:SpiralRecons}
\end{figure}

\begin{figure}[h]
\centering
\includegraphics[width=\linewidth]{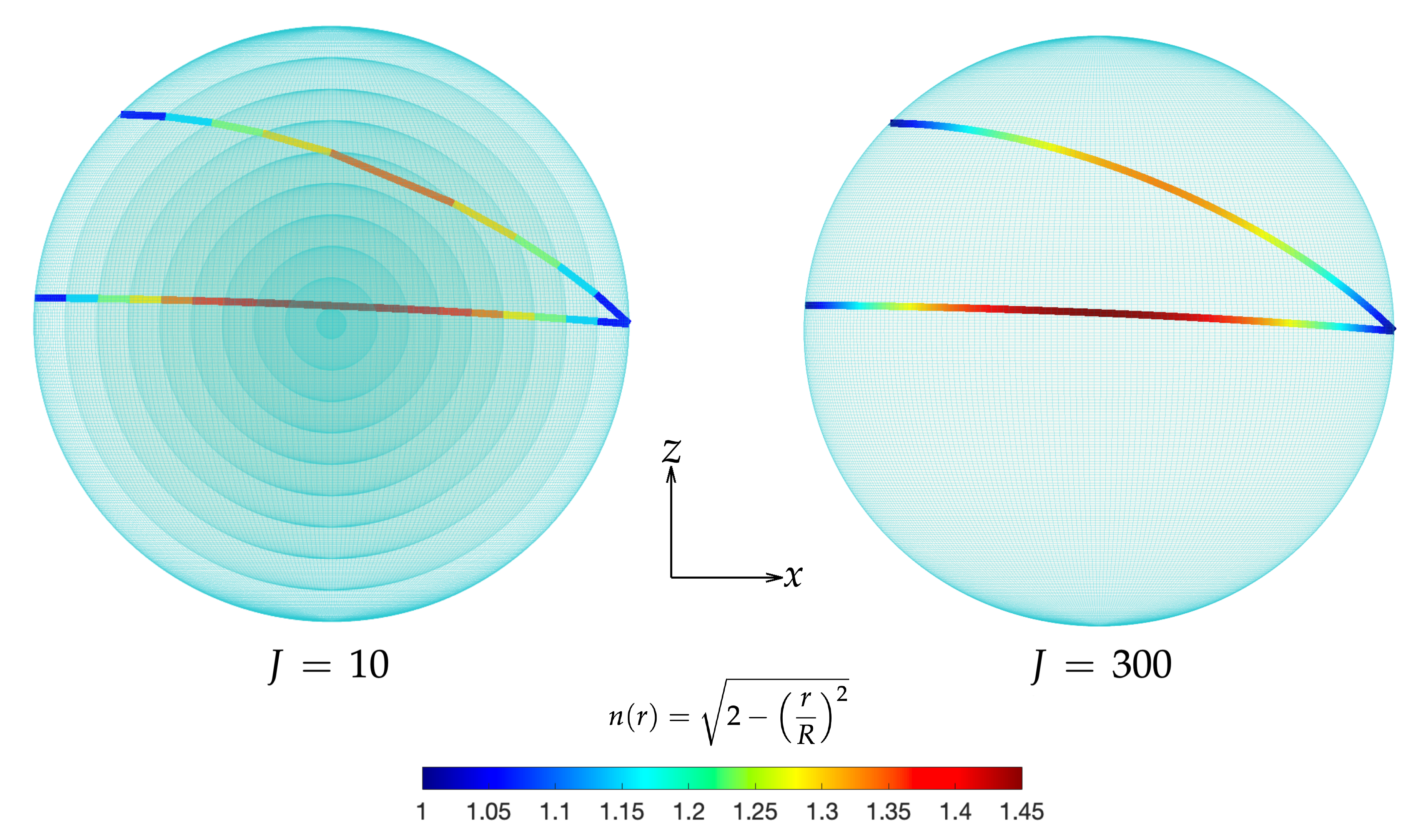}
\caption{Effect of ray position and interface number on the Luneburg reconstruction. Two prescribed trajectories are shown for $J=10$ (left) and $J=300$ (right) concentric spherical interfaces. From the left image we can see that the ray closer to the optical axis crosses more interfaces and samples a wider radial interval. Increasing $J$ refines the reconstruction along each trajectory without extending its accessible radial interval. Colors indicate the recovered refractive index.}
    \label{fig:TwoRays}
\end{figure}

For this reconstruction, we use $J=300$ concentric spherical interfaces as the prescribed family of surfaces, consistent with the known spherical symmetry of the Luneburg lens. The reconstruction is initialized at the surface of the spherical lens with a refractive index of $n_1=1$, since the Luneburg refractive index matches that of the surrounding medium at the lens boundary and is normalized to this value. Figure \ref{fig:SpiralRecons}(a) shows the refractive index reconstructed along each prescribed trajectory. 
As seen in Fig. \ref{fig:SpiralRecons}(b), the radial extent of the reconstruction depends on how closely the trajectory approaches the lens center. Rays entering closer to the optical axis pass through more concentric interfaces, recovering the refractive index over a wider radial interval. Peripheral rays intersect fewer interfaces and recover only the outer portion of the refractive index profile. Both cases are illustrated in the left panel of Fig. \ref{fig:TwoRays}.

For a given trajectory, the reconstruction also depends on the number of interfaces $J$, as shown in Fig.~\ref{fig:TwoRays}. Increasing $J$ refines the radial sampling within the region traversed by the ray, without extending the radial interval accessible to that trajectory. Thus, the trajectory determines the extent of the reconstructed region, whereas the interface density controls the discretization used to recover the refractive index profile.

For the on-axis ray in the Luneburg lens, incidence and transmission are normal to the concentric spherical interfaces, so both angles are zero and Eq.~\eqref{eq:reconstruction} becomes indeterminate. This ray crosses the interfaces without angular deflection regardless of the refractive index values, which therefore cannot be recovered from its trajectory alone. The on-axis ray is consequently excluded from the reconstruction accuracy assessment.

Since rays passing closer to the lens center recover the refractive index over a wider radial interval, we select the nonaxial ray whose entry point has the smallest normalized radial distance from the optical axis, $0.0872$, for the quantitative accuracy assessment. The reconstruction error is evaluated using the root mean square error,
\begin{equation}
\label{eq:rmse}
\mathrm{RMSE}=
\sqrt{\frac{1}{M}\sum_{m=1}^{M}
\left[n_m-n(r_m)\right]^2},
\end{equation}
where $M=2J-2$ is the maximum value of the reconstruction sample index $m$, $n_m$ is the recovered refractive index, and $n(r_m)$ is the analytical refractive index evaluated at the corresponding radial position $r_m$ using Eq.~\eqref{eq:luneburg}.
\begin{figure}[h]
\centering
\includegraphics[width=\linewidth]{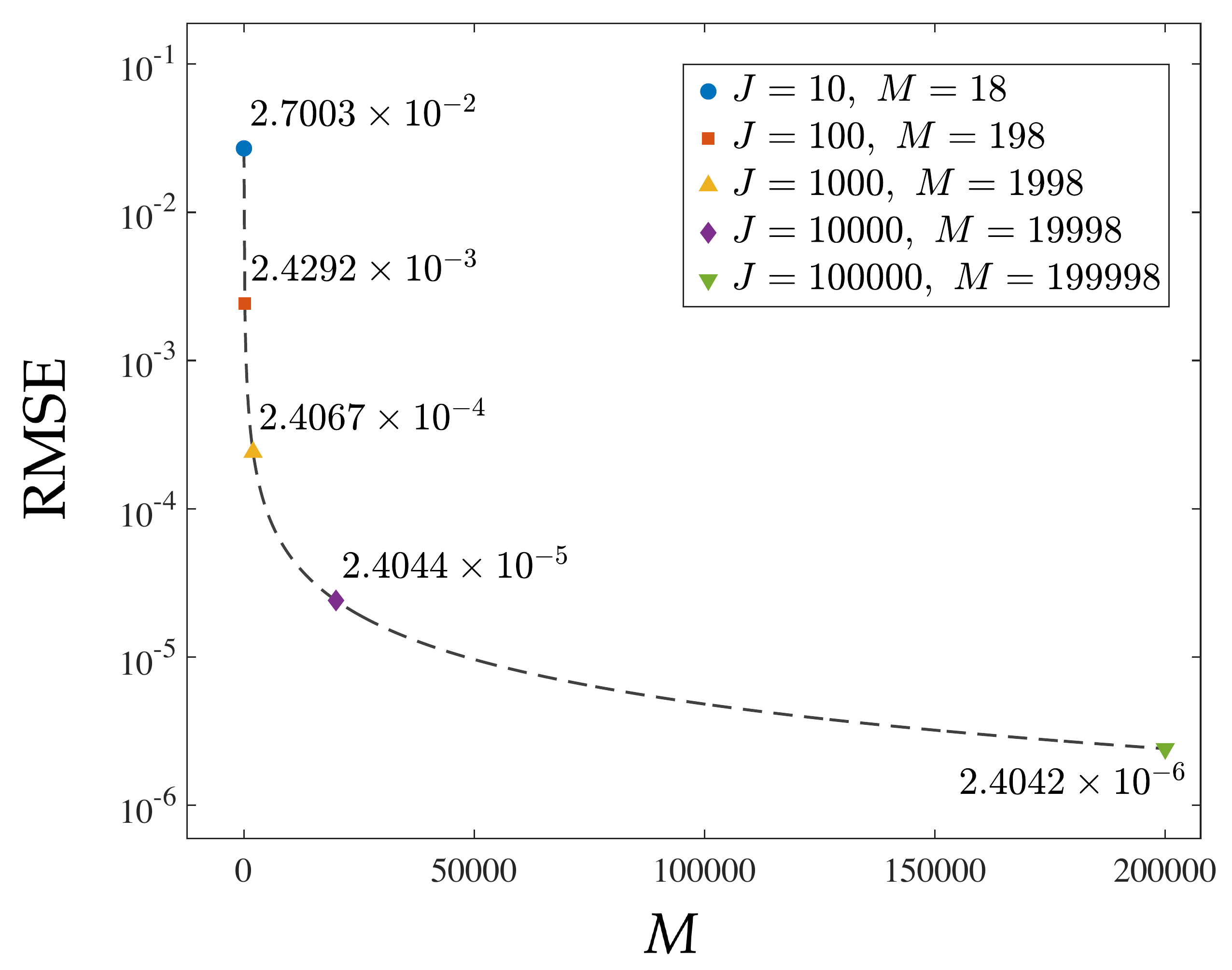}
\caption{Root mean square error of the reconstructed Luneburg refractive index profile for the nonaxial ray with normalized incident height $0.0872$. Results are shown for $J=10$ to $10^5$ reconstruction interfaces, with the corresponding numbers of reconstructed samples $M$ indicated in the legend. The error decreases from $2.700\times10^{-2}$ to $2.404\times10^{-6}$, consistent with approximately first-order convergence for this test.}
    \label{fig:RMSE}
\end{figure}

Figure~\ref{fig:RMSE} shows the RMSE as a function of $M$ for increasing numbers of reconstruction interfaces $J$. The error decreases monotonically from $2.700\times10^{-2}$ for $J=10$ ($M=18$) to $2.404\times10^{-6}$ for $J=10^5$ ($M=199998$), a reduction of slightly more than four orders of magnitude. For the finer discretizations, each tenfold increase in $J$ produces an approximately tenfold reduction in the RMSE, consistent with first-order convergence for this test. No error plateau is observed over the evaluated range, indicating that refinement continues to improve the reconstruction of the analytical refractive index profile.



\section{Freeform GRIN reconstruction}

\begin{figure}[h!]
\centering
\includegraphics[width=0.9\linewidth]{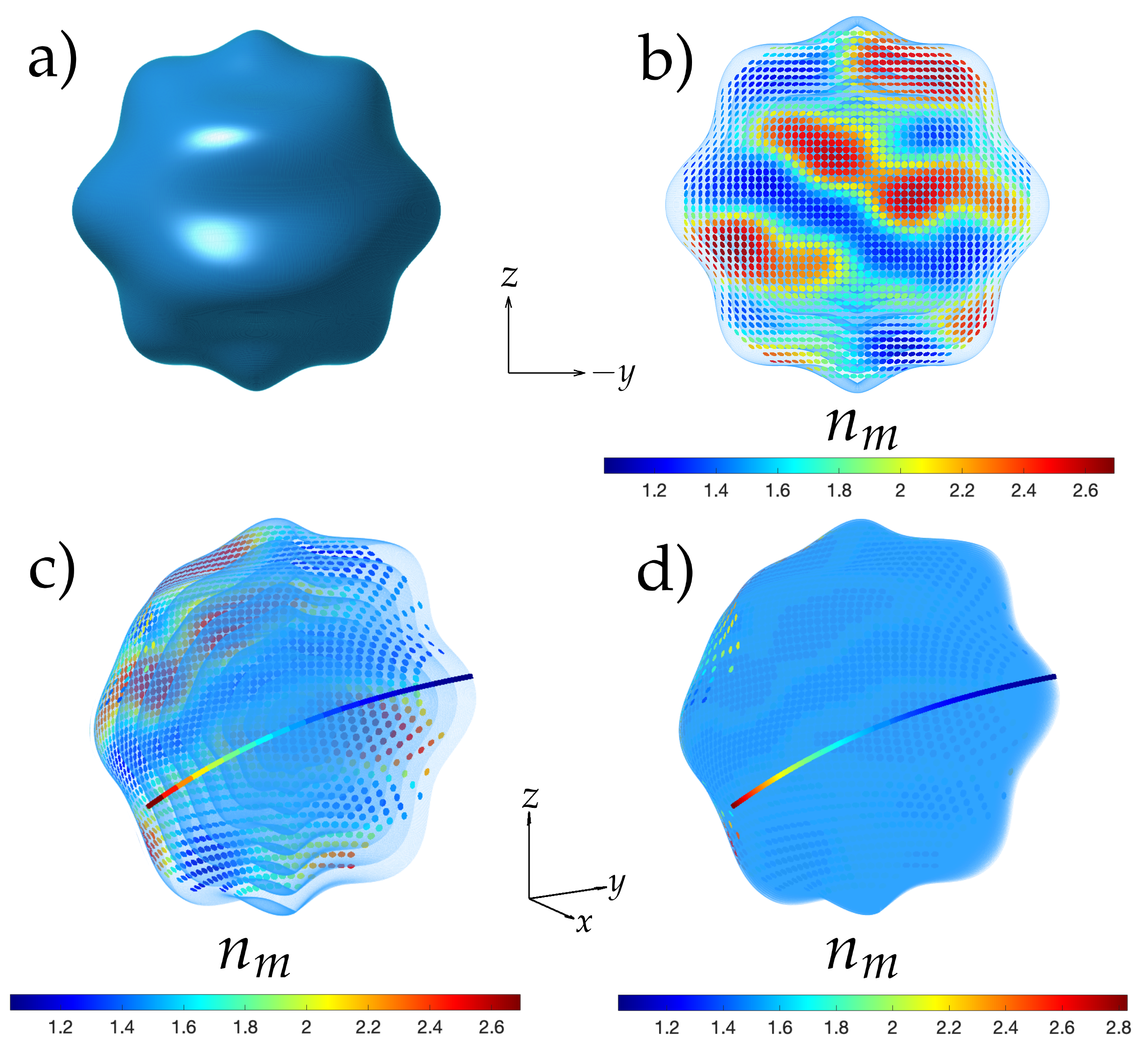}
\caption{Freeform GRIN reconstruction using the nested interfaces defined by Eq.~\eqref{eq:interfaces} with $a=0.075$, $k_1=7$, and $k_2=2$. (a) Outermost interface viewed along the $x$ axis. (b) Local refractive index assignment on this interface using 1876 parallel incident rays and randomly prescribed admissible transmission directions. (c,d) Reconstruction along the selected trajectory $\boldsymbol{\gamma}(s)$ for $J=10$ and $J=100$, respectively. The surrounding refractive index is $n_0=1$, and $\delta=0.04$ controls the transverse extent of the local assignments. Colors represent the recovered refractive index.}
    \label{fig:FreeformSurfaces}
\end{figure}

Having assessed the reconstruction accuracy using a Luneburg lens, we now consider a prescribed trajectory for which the refractive index distribution is unknown. We apply the method using two families of reconstruction interfaces with different geometries to examine how this choice affects the recovered refractive index distribution and its suitability for physical realization. These interfaces are not constrained to be iso-indicial surfaces, so the refractive index may vary across each interface. We first consider the family of interfaces defined by
\begin{equation}
\label{eq:interfaces}
\mathbf{\Sigma}_j(\theta,\phi)=
\rho_j\left[1+a\sin(k_1\theta)\cos(k_2\phi)\right]
\begin{bmatrix}
\sin\theta\cos\phi\\
\sin\theta\sin\phi\\
\cos\theta
\end{bmatrix},
\end{equation}
where $\theta\in[0,\pi]$, $\phi\in[0,2\pi)$, $a=0.075$, $k_1=7$, and $k_2=2$. The integer $j = 1, \dots, J$ labels the interfaces, while $\rho_j \in (0, 1]$ specifies their scale
relative to the outermost surface, defined by $\rho_1 = 1$. The $\rho_j$ scale factors therefore
define a family of nested interfaces bounded by this surface. Figure~\ref{fig:FreeformSurfaces}(a) shows the outermost interface viewed along the $x$ axis (projected onto the $y$–$z$ plane), revealing the external shape of the medium. We distribute the incident rays across the full transverse extent of the medium, ensuring that their intersections sample the illuminated portion of the outermost interface.

A set of 1876 parallel rays, representing an incident plane wavefront propagating along the $\hat{x}$ direction, is distributed across the full transverse extent of the medium shown in Fig.~\ref{fig:FreeformSurfaces}(a). The surrounding medium has a refractive index of $n_0=1$, which provides the initial condition for the reconstruction. To demonstrate the local recovery of the refractive index on a single interface, we use the outermost surface as an example and randomly prescribe a transmitted direction at each ray–interface intersection, subject to the local refraction conditions. The corresponding refractive index values are calculated using Eq.~\eqref{eq:reconstruction} and assigned to local regions whose transverse extent is controlled by $\delta=0.04$. Figure~\ref{fig:FreeformSurfaces}(b) shows the resulting refractive index distribution on this interface. 

\begin{figure}[h!]
\centering
\includegraphics[width=\linewidth]{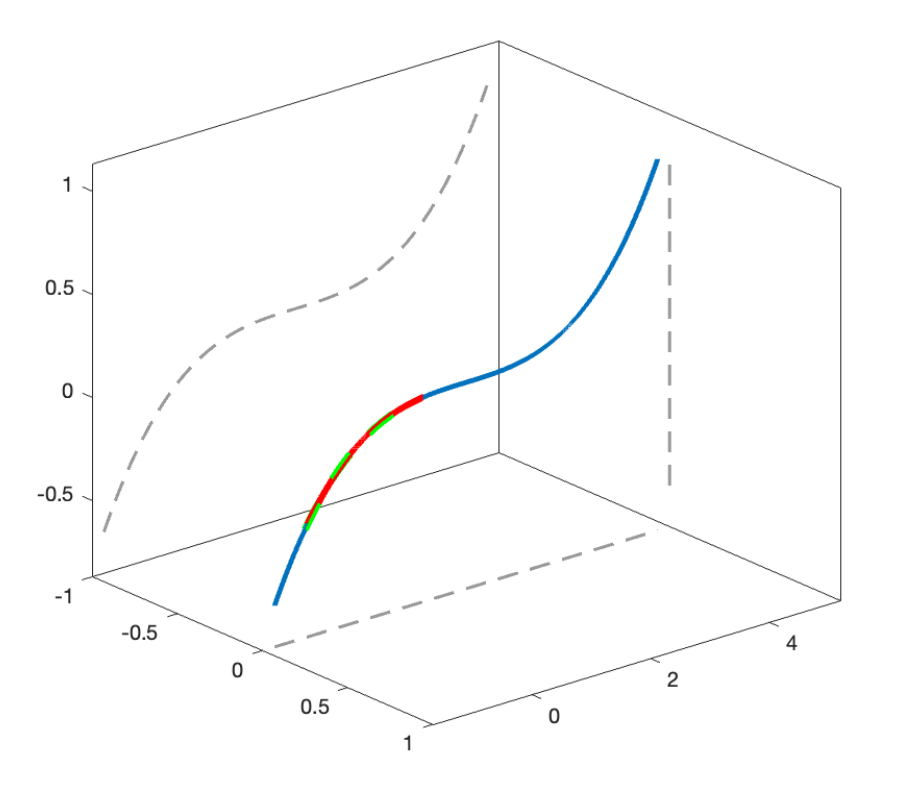}
\caption{Planar trajectory defined by Eq. \eqref{eq:prescribed_trajectory}, evaluated over $s\in[0,1]$. The red solid and green dashed segments indicate the portions used for reconstruction with the first interface family, $s\in[0.0872,0.3868]$, and the concentric spherical interfaces, $s\in[0.0794,0.3787]$, respectively. Blue portions show the remainder of the curve. Gray dashed lines show projections onto the coordinate planes.}
    \label{fig:trajectory_planar}
\end{figure}

The reconstruction presented so far is limited to a single interface. To illustrate how the reconstruction proceeds through successive interfaces, we consider a single ray with the prescribed trajectory
\begin{equation}
\label{eq:prescribed_trajectory}
\boldsymbol{\gamma}(s)=
\begin{bmatrix}
0\\
-1.45+6.45s\\
0.13+0.8(2s-1)^3
\end{bmatrix},
\qquad s\in[0.0872,0.3868].
\end{equation}
Figure~\ref{fig:trajectory_planar} shows the prescribed trajectory over $s\in[0,1]$, with the interval $s\in[0.0872,0.3868]$ used for reconstruction highlighted in red. The ray is initially parallel to the $y$ axis and enters the medium at $\mathbf{P}_1=\boldsymbol{\gamma}(0.0872)\approx(0,-0.8875,-0.3202)$.

Using this prescribed trajectory, we reconstruct the refractive index distribution for two discretizations of the same interface family: $J=10$ and $J=100$, shown in Figs.~\ref{fig:FreeformSurfaces}(c) and \ref{fig:FreeformSurfaces}(d), respectively. In both cases, the refractive index is recovered successively along the ray path and assigned to local regions around it. Increasing the number of interfaces refines the piecewise-homogeneous representation of the medium, allowing the propagation segments to follow $\boldsymbol{\gamma}$ more closely. The recovered refractive index values range from $1.0360$ to $2.6927$ for $J=10$ and from $1.0272$ to $2.8294$ for $J=100$. The colors across each interface represent local refractive index assignments within regions of transverse extent controlled by $\delta$, without imposing a uniform refractive index over the entire interface.

\begin{figure}[h!]
\centering
\includegraphics[width=\linewidth]{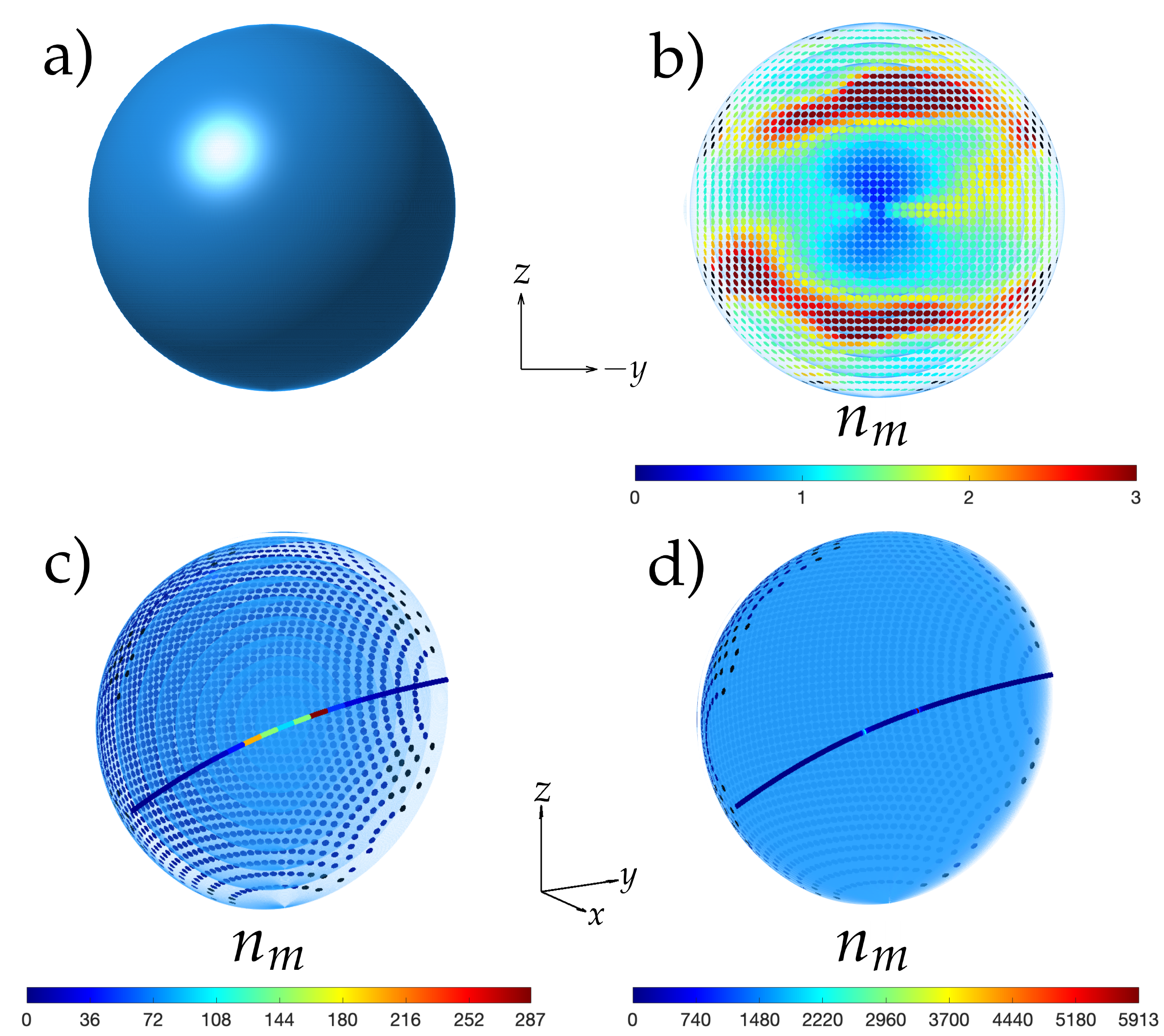}
\caption{Freeform GRIN reconstruction using concentric spherical interfaces, obtained by setting $a=0$ in Eq.~\eqref{eq:interfaces}. (a) Outermost interface. (b) Local refractive index assignment using the same incident rays and prescribed transmitted directions as in Fig.~\ref{fig:FreeformSurfaces}(b), with $n_0=1$ and $\delta=0.04$. Colored points represent recovered refractive index values, including values below unity. Black points mark locations where the prescribed directions do not satisfy the local refraction conditions and no refractive index value is assigned. In panel (b), the color scale is capped at $n=3$ to make variations at lower refractive index values visible. Values above this limit are displayed using the same upper-end color, without modifying the recovered refractive indices. (c,d) Reconstruction using the same trajectory function $\boldsymbol{\gamma}(s)$ for $J=10$ and $J=100$, respectively. The recovered refractive index values range from $2.9725$ to $286.8784$ in (c) and from $2.6753$ to $5.9129\times10^3$ in (d). Panels (b), (c), and (d) use separate color scales.}
    \label{fig:ray_tracing_esfera}
\end{figure}

To illustrate the importance of interface selection for physical realization, we deliberately consider a second geometry that yields refractive indices below unity on the outermost interface and very large values along the prescribed trajectory. This example highlights the need to choose the reconstruction interfaces together with the prescribed ray directions to obtain refractive indices within a desired range. We set $a=0$ in Eq.~\eqref{eq:interfaces}, obtaining a family of concentric spherical interfaces. Figure~\ref{fig:ray_tracing_esfera}(a) shows the outermost interface. To reconstruct the refractive index distribution on this surface, we use the same set of 1876 parallel incident rays representing a plane wavefront and the same prescribed transmitted directions as in the previous example, retaining $n_0=1$ and $\delta=0.04$. Although the incident and transmitted directions remain unchanged, the new interface normals modify both the incidence and transmission angles.


The local refraction conditions must therefore be checked again at each ray--interface intersection. Changing the interface geometry can cause the local normal to lie outside the plane defined by the prescribed incident and transmitted directions, violating the coplanarity condition. At these points, no scalar refractive index can reproduce the prescribed transmitted direction under Snell's law for isotropic media. These points are shown in black in Fig.~\ref{fig:ray_tracing_esfera}(b), indicating that no refractive index value is assigned there. Wherever the local refraction conditions are satisfied, the refractive index is recovered using Eq.~\eqref{eq:reconstruction}. For visualization, the color scale in Fig.~\ref{fig:ray_tracing_esfera}(b) is capped at $n=3$, with all higher values represented by the same upper-end color. This prevents the largest recovered indices from obscuring variations at lower values and does not alter the calculated refractive indices.

The recovered values include refractive indices below unity. With $n_0=1$, these values arise wherever the transmission angle exceeds the incidence angle for an admissible refraction event. Thus, changing the interface geometry while retaining the prescribed ray directions affects both the recovered refractive index values and the locations where the reconstruction is admissible. This result highlights the need to consider the interface geometry and the prescribed ray directions together when seeking a distribution within a desired refractive index range. The variation of the refractive index across this spherical interface also emphasizes that the geometry of the reconstruction surfaces does not impose spherical symmetry on the recovered distribution.

We then retain the prescribed trajectory function $\boldsymbol{\gamma}(s)$ to reconstruct the refractive index inside the medium. The change in the outer boundary modifies the portion of this trajectory contained within the medium, which now corresponds to $s\in[0.0794,0.3787]$, with the incident ray entering at $\mathbf{P}_1=\boldsymbol{\gamma}(0.0794)\approx(0,-0.9379,-0.3462)$. Thus, the trajectory function is unchanged, but its entry and exit points are determined by the new boundary. The corresponding interval, $s\in[0.0794,0.3787]$, is indicated by the green dashed segment in Fig.~\ref{fig:trajectory_planar}.

Figures~\ref{fig:ray_tracing_esfera}(c) and \ref{fig:ray_tracing_esfera}(d) show the reconstruction for $J=10$ and $J=100$, respectively. The recovered refractive index values range from $2.9725$ to $286.8784$ for $J=10$ and from $2.6753$ to $5.9129\times10^3$ for $J=100$, substantially exceeding those obtained with the previous interface family. Changing the interface geometry modifies the intersection points and local normals along the prescribed trajectory, thereby changing the refraction angles and the refractive index ratios required at successive intersections. The resulting distributions therefore differ even though they are reconstructed from the same trajectory function. 

The large indices obtained with spherical interfaces arise at reconstruction steps where the transmitted segment is nearly aligned with the local normal. At these intersections, $\sin\theta_{t,m}$ becomes small without a proportional decrease in $\sin\theta_{i,m}$, producing a large refractive index ratio in Eq.~\eqref{eq:reconstruction}. Changing $J$ modifies the intersection positions and the directions of the adjacent segments, and therefore how closely a transmitted segment aligns with the normal. This accounts for the substantial increase in the maximum recovered index between the two spherical discretizations. Since the reconstruction is recursive, these local ratios also affect subsequent index values. Thus, both the interface geometry and its discretization influence the refractive indices required to reproduce the prescribed trajectory.

\begin{figure}[h!]
\centering
\includegraphics[width=\linewidth]{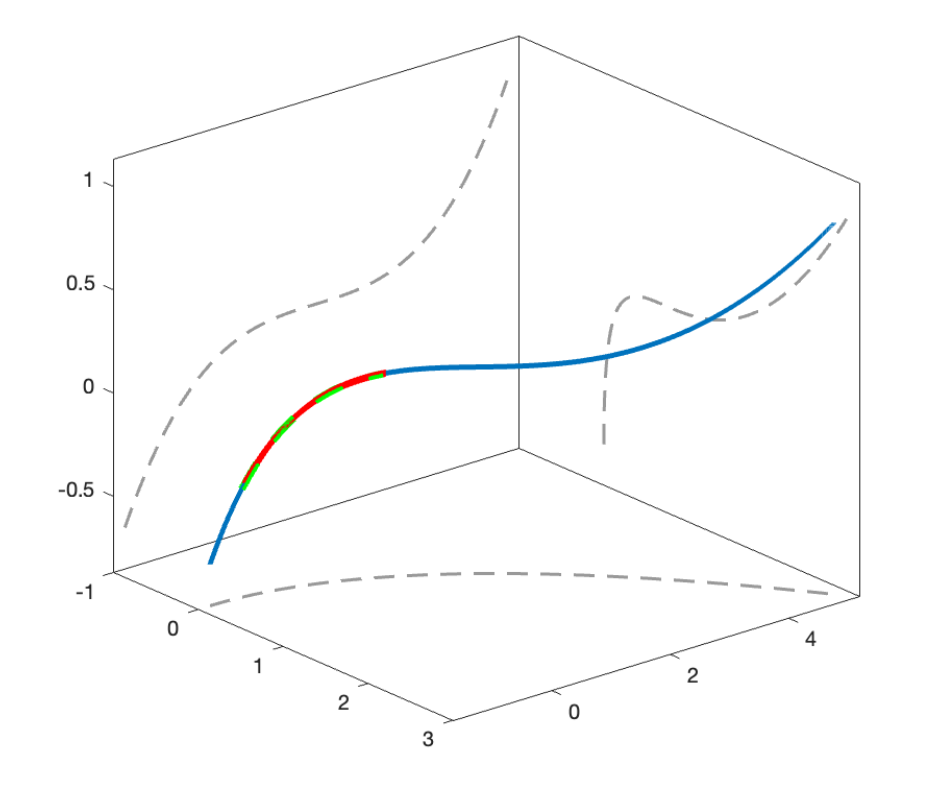}
\caption{Nonplanar trajectory defined by Eq. \eqref{eq:prescribed_trajectory_2}, evaluated over $s\in[0,1]$. The red solid and green dashed segments indicate the portions used for reconstruction with the first interface family, $s\in[0.0872,0.3724]$, and the concentric spherical interfaces, $s\in[0.0794,0.3669]$, respectively. Blue portions show the remainder of the curve. Gray dashed lines show projections onto the coordinate planes.}
    \label{fig:trajectory_nonplanar}
\end{figure}

The trajectory considered so far lies in the $y-z$ plane. To extend these examples to a nonplanar path, we introduce a quadratic term in its $x$ coordinate while retaining the expressions for $y(s)$ and $z(s)$ in Eq.~\eqref{eq:prescribed_trajectory}:
\begin{equation}
\label{eq:prescribed_trajectory_2}
\boldsymbol{\gamma}_{\mathrm{3D}}(s)=
\begin{bmatrix}
2.85s^{2}\\
-1.45+6.45s\\
0.13+0.8(2s-1)^{3}
\end{bmatrix}.
\end{equation}
This trajectory is not contained in a single plane and is shown in Fig.~\ref{fig:trajectory_nonplanar} over $s\in[0,1]$. The intervals used for reconstruction are $s\in[0.0872,0.3724]$ for the first interface family and $s\in[0.0794,0.3669]$ for the concentric spherical interfaces, shown as red solid and green dashed segments, respectively.

 We apply the same reconstruction procedure using both interface families and both discretizations, $J=10$ and $J=100$. For each geometry, the ray–interface intersections and adjacent segment directions are recalculated, with the outer boundary determining the portion of the trajectory inside the medium.
\begin{figure}[h!]
\centering
\includegraphics[width=0.9\linewidth]{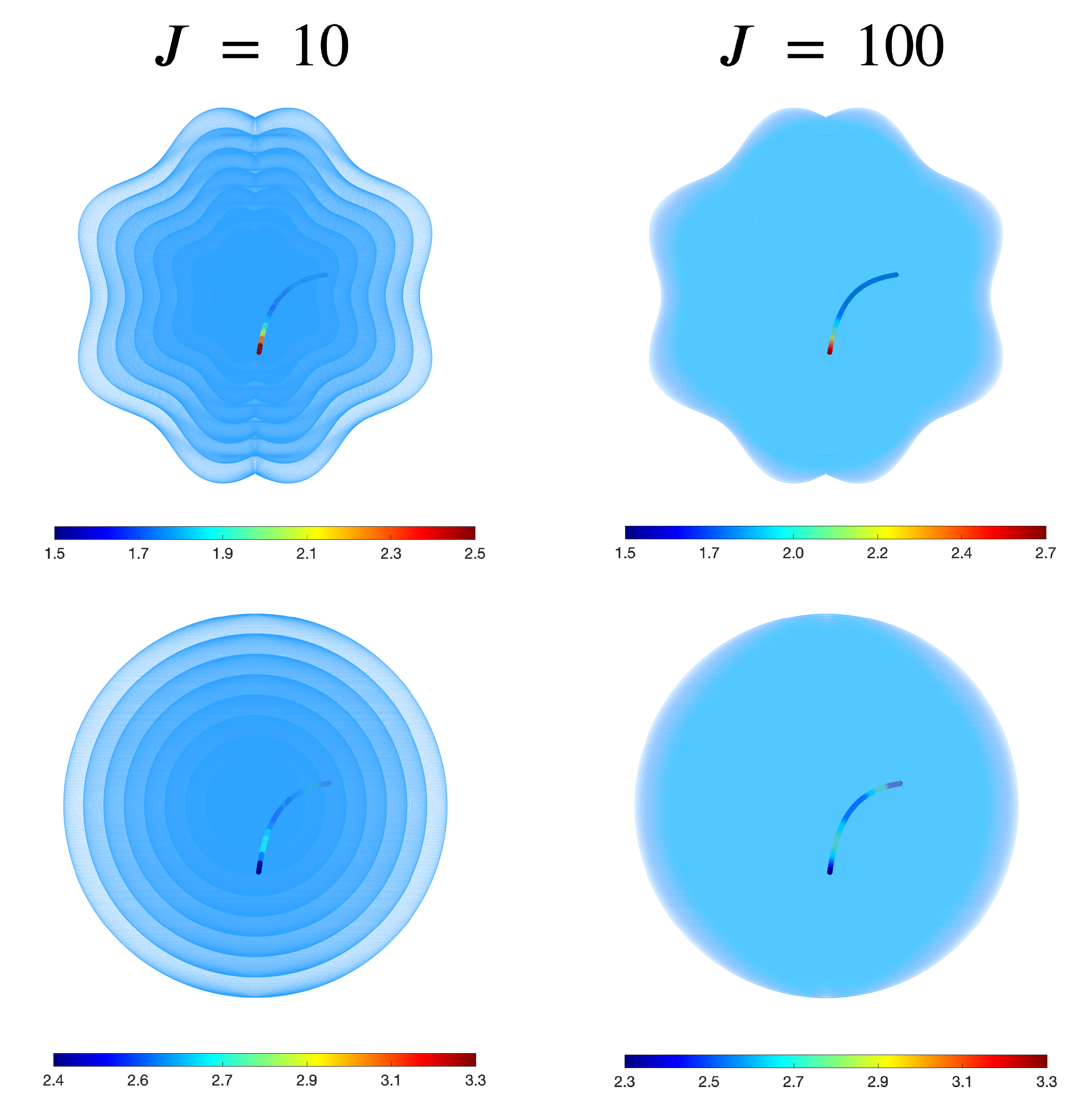}
\caption{Local reconstruction of freeform GRIN distributions along the prescribed nonplanar trajectory using the interfaces defined by Eq. Eq.~\eqref{eq:interfaces}, with $a=0.075$ (top row) and $a=0$ (bottom row). The left and right columns correspond to $J=10$ and $J=100$, respectively. Each panel has an independent color scale representing only the refractive index values reconstructed along the ray path. The translucent blue surfaces show the reconstruction interfaces; their color does not represent a refractive index value or imply a uniform index within the intervening layers.}
    \label{fig:FueraDelPlano}
\end{figure}

Figure \ref{fig:FueraDelPlano} shows the local refractive index distributions reconstructed along this nonplanar trajectory. For the interfaces defined by Eq. \eqref{eq:interfaces} with $a=0.075$, the recovered indices range from $1.5105$ to $2.5373$ for $J=10$ and from $1.5027$ to $2.6692$ for $J=100$. For the concentric spherical interfaces, obtained with $a=0$, the corresponding ranges are $2.3989$–$3.2601$ and $2.2797$–$3.3001$, respectively. Unlike the large values obtained for the previous planar trajectory, the same spherical interface family yields substantially lower refractive indices for this nonplanar path. This comparison shows that the recovered index range depends on the prescribed trajectory as well as the interface geometry.

\section{Refractive index compatibility}
The preceding examples reconstruct the refractive index locally along individual prescribed trajectories. Combining several such reconstructions into a single optical medium requires their local index assignments to be compatible. Although different interface geometries may yield different distributions for the same trajectory, the refractive index within any individual isotropic medium must be a single-valued function of position at a fixed wavelength. This requirement ensures consistency within a reconstructed medium without implying uniqueness of the inverse solution.

Consider two prescribed trajectories that intersect at a point $\mathbf{P}^{\ast}$ within a locally homogeneous region. Their independently recovered refractive index values must satisfy
\begin{equation}
\label{eq:compatibility}
n^{(a)}(\mathbf{P}^{\ast})=
n^{(b)}(\mathbf{P}^{\ast}),
\end{equation}
where the superscripts identify the two reconstructions. The trajectories may cross with different propagation directions, but they cannot require different refractive index values at the same position. Conflicting assignments prevent the reconstructions from being combined directly under the selected conditions and require revising the interface geometry, the prescribed trajectories, or the initial conditions.

Since each recovered value is assigned to a finite local volume, compatibility must also hold wherever reconstructed regions overlap, even when the ray trajectories themselves do not intersect. Denoting the regions associated with two reconstructions by $V_a$ and $V_b$, this condition becomes
\begin{equation}
\label{eq:volume_compatibility}
n^{(a)}(\mathbf{r})=n^{(b)}(\mathbf{r}),
\qquad \mathbf{r}\in V_a\cap V_b.
\end{equation}
The parameter $\delta$ therefore controls both the transverse extent of the reconstruction and the overlap between local assignments. Reducing $\delta$ may eliminate overlaps between spatially separated trajectories, but it cannot resolve conflicting refractive index requirements at an actual ray intersection.

These conditions ensure consistency of the local refractive index assignments. 
Practical designs may also impose an admissible range, $n_{\min}\leq n(\mathbf{r})\leq n_{\max}$, determined by the available materials and operating wavelength. These bounds must be incorporated into the reconstruction constraints, since simply replacing recovered values outside this range generally changes the prescribed ray trajectories.

\section{Discussion}

The results illustrate how the choice of reconstruction interfaces provides geometrical freedom in recovering a refractive index distribution from a prescribed ray trajectory. Changing from the first interface geometry to concentric spherical interfaces while retaining the same trajectory function produces different local refractive index assignments. This dependence reflects the nonuniqueness of the inverse problem and makes the interface geometry a design choice within the reconstruction procedure. Furthermore, spherical reconstruction interfaces do not necessarily yield a spherically symmetric distribution, since the recovered refractive index may vary across each spherical layer.

This geometrical freedom is subject to the local refraction conditions. When the incident and transmitted directions are retained, changing the interface normals modifies their relative orientation and can affect whether an admissible reconstruction exists at a given intersection. The black points in Fig.~\ref{fig:ray_tracing_esfera}(b) identify locations where these conditions are not satisfied and no refractive index is assigned. At admissible intersections, the change in geometry can also produce refractive index values below unity, as observed for the spherical interface. These results show that the interface geometry and the prescribed ray directions must be considered together, both to satisfy the refraction conditions and to obtain values within a desired refractive index range.

The Luneburg benchmark demonstrates that refining the reconstruction interfaces improves agreement with the analytical refractive index profile, with approximately first-order convergence over the finer discretizations examined. In that test, the interfaces follow the known spherical symmetry of the reference distribution. For the two interface geometries considered in Section~4, where no reference distribution is prescribed, increasing the number of interfaces changes the sampling of the trajectory and the local refraction angles used in the reconstruction. A finer discretization therefore does not by itself ensure that the recovered indices remain within a range suitable for a particular application. The large values obtained along the selected trajectory with spherical interfaces emphasize the need to consider discretization together with interface geometry and refractive index constraints.

The method reconstructs local regions around prescribed trajectories. Combining these regions into a single optical medium requires consistent refractive index assignments wherever they overlap, as discussed in Section~5. Practical applications may also require the recovered indices to lie within an admissible range. These bounds constrain the choice of reconstruction interfaces and prescribed trajectories, since simply replacing values outside the desired range generally changes the refraction conditions and, consequently, the resulting ray paths.

The nonplanar example further illustrates this joint dependence: the same spherical interface family that yields very large refractive indices for the planar trajectory produces substantially lower values when the prescribed path is changed. Thus, the suitability of an interface geometry for physical realization must be assessed in relation to the trajectory it is intended to reproduce.

\section{Conclusions}

We have presented a point-wise method for reconstructing gradient-index media from prescribed ray trajectories through the inverse application of local refraction. The formulation applies to three-dimensional geometries, with two-dimensional reconstruction as a particular case, and requires neither geometrical symmetry nor prior knowledge of the iso-indicial surfaces. A key advantage is its straightforward implementation: each reconstruction step uses the adjacent ray directions and the local interface normal to recover the refractive index through a simple algebraic relation, without integrating the ray differential equation. The Luneburg lens benchmark demonstrates agreement with the analytical refractive index profile, with approximately first-order convergence for the finer discretizations examined.

The comparison between the two interface geometries shows that different refractive index distributions can be recovered from the same prescribed trajectory function. The reconstruction interfaces therefore provide an additional degree of freedom for optical design, subject to the local refraction conditions. Their geometry, the prescribed trajectories, and the discretization must be considered together when seeking a desired refractive index range. For multiple trajectories, the local assignments must also be consistent wherever the reconstructed regions overlap. Within these constraints, the method provides a readily implementable framework for the local reconstruction of symmetric and freeform GRIN media.


\section*{CRediT authorship contribution statement}

\textbf{V. Ch\'{a}vez-Islas:} Formal analysis, Investigation, Methodology, Software, Visualization, Writing - original draft. \textbf{F. H. Maldonado-Villamizar:} Formal analysis, Investigation, Methodology, Visualization, Writing - original draft. \textbf{E. Esp\'{i}ndola-Ramos:} Formal analysis, Investigation, Methodology, Visualization, Writing - original draft. \textbf{S. Ch\'{a}vez-
Cerda:} Investigation, Methodology, Supervision, Validation, Visualization, Writing - original draft. \textbf{J. E. G\'{o}mez-Correa:} Conceptualization, Formal analysis, Investigation, Methodology, Project administration, Supervision, Validation, Visualization, Writing -original draft .

\section*{Acknowledgments}
The first and third authors thank the Secretar\'{i}a de Ciencia, Humanidades, Tecnolog\'{i}a e Innovaci\'{o}n (SECIHTI) for the support provided through the awarded scholarships (VCI: 1348421; EER: 640385). FHMV also acknowledges support from SECIHTI through project IIxM 7309 (CIR/072/2024). 

\section*{Data availability}
No data was used for the research described in the article.

\section*{Declaration of competing interest}
The authors declare that they have no known competing financial interests or personal relationships that could have appeared to influence the work reported in this paper.

\bibliography{OLTRef}

@book{GomezLibro, 
 author={C. G\'{o}mez-Reino and M. V. P\'{e}rez and C. Bao},
title={Gradient-Index Optics: Fundamentals and Applications},  
publisher={Springer}, 
place={New York},
year={2002}
}

@book{Luneburg, 
 author={R. K. Luneburg},
title={Mathematical Theory of Optics},  
publisher={University of California}, 
place={Los Angeles California},
year={1964}
}

@book{Lakshminarayanan, 
author={V. Lakshminarayanan and A. K. Ghatak and K. Thyagarajan}, 
title={Lagrangian optics},  
publisher={Springer}, 
place={USA}, 
year={2002}
}

@article{Lippman:21,
author = {David H. Lippman and Nicholas S. Kochan and Tianyi Yang and Greg R. Schmidt and Julie L. Bentley and Duncan T. Moore},
journal = {Opt. Express},
number = {22},
pages = {36997--37012},
publisher = {Optica Publishing Group},
title = {Freeform gradient-index media: a new frontier in freeform optics},
volume = {29},
month = {Oct},
year = {2021},
url = {https://opg.optica.org/oe/abstract.cfm?URI=oe-29-22-36997},
doi = {10.1364/OE.443427},
}

@article{Manipulating,
author = {Wanguo Liu and Hai Hu and Fenghua Liu and Hua Zhao},
journal = {Opt. Express},
number = {4},
pages = {4714--4726},
publisher = {OSA},
title = {Manipulating light trace in a gradient-refractive-index medium: a {Lagrangian} optics method},
volume = {27},
month = {Feb},
year = {2019},
url = {http://www.opticsexpress.org/abstract.cfm?URI=oe-27-4-4714},
doi = {10.1364/OE.27.004714},
}

@article{Nemoto:79,
author = {S. Nemoto and T. Makimoto},
journal = {J. Opt. Soc. Am.},
number = {3},
pages = {450--454},
publisher = {Optica Publishing Group},
title = {Refractive-index distribution for a prescribed ray path},
volume = {69},
month = {Mar},
year = {1979},
url = {https://opg.optica.org/abstract.cfm?URI=josa-69-3-450},
doi = {10.1364/JOSA.69.000450},
}

@article{Minano:85,
author = {J. C. {Mi\~{n}ano}},
journal = {J. Opt. Soc. Am. A},
number = {11},
pages = {1821--1825},
publisher = {Optica Publishing Group},
title = {Refractive-index distribution in two-dimensional geometry for a given one-parameter manifold of rays},
volume = {2},
month = {Nov},
year = {1985},
url = {https://opg.optica.org/josaa/abstract.cfm?URI=josaa-2-11-1821},
doi = {10.1364/JOSAA.2.001821},
}

@article{Borghero_2005,
doi = {10.1088/0305-4470/38/1/012},
url = {https://dx.doi.org/10.1088/0305-4470/38/1/012},
year = {2004},
month = {dec},
publisher = {},
volume = {38},
number = {1},
pages = {175},
author = {F Borghero and G Bozis},
title = {A two-dimensional inverse problem of geometrical optics},
journal = {Journal of Physics A: Mathematical and General}
}

@article{Borghero:06,
author = {Francesco Borghero and George Bozis},
journal = {J. Opt. Soc. Am. A},
number = {12},
pages = {3133--3138},
publisher = {Optica Publishing Group},
title = {Two solvable problems of planar geometrical optics},
volume = {23},
month = {Dec},
year = {2006},
url = {https://opg.optica.org/josaa/abstract.cfm?URI=josaa-23-12-3133},
doi = {10.1364/JOSAA.23.003133},
}

@article{Borghero:11,
author = {Francesco Borghero and Thomas Kotoulas},
journal = {J. Opt. Soc. Am. A},
number = {2},
pages = {278--283},
publisher = {Optica Publishing Group},
title = {Refractive-index distributions generating as light rays a given family of curves lying on a surface},
volume = {28},
month = {Feb},
year = {2011},
url = {https://opg.optica.org/josaa/abstract.cfm?URI=josaa-28-2-278},
doi = {10.1364/JOSAA.28.000278},
}

@article{Borghero:16,
author = {Francesco Borghero and Francesco Demontis},
journal = {J. Opt. Soc. Am. A},
number = {9},
pages = {1710--1722},
publisher = {Optica Publishing Group},
title = {Three-dimensional inverse problem of geometrical optics: a mathematical comparison between {Fermat's} principle and the eikonal equation},
volume = {33},
month = {Sep},
year = {2016},
url = {https://opg.optica.org/josaa/abstract.cfm?URI=josaa-33-9-1710},
doi = {10.1364/JOSAA.33.001710},
}

@article{GRINMapping,
author = {Hiroshi Ohno and Takashi Usui},
journal = {OSA Continuum},
number = {10},
pages = {2543--2551},
publisher = {Optica Publishing Group},
title = {Neural network gradient-index mapping},
volume = {4},
month = {Oct},
year = {2021},
url = {https://opg.optica.org/osac/abstract.cfm?URI=osac-4-10-2543},
doi = {10.1364/OSAC.437395},
}

@article{GRINMapping2,
author = {Hiroshi Ohno and Takashi Usui},
journal = {Optical Review},
number = {},
pages = {1--6},
publisher = {},
title = {Design method of gradient-index optics in homogeneous medium using neural network gradient-index mapping},
volume = {},
month = {},
year = {2023},
url = {},
doi = {10.1007/s10043-023-00803-1},
}

@article{Gomez-Correa:22,
author = {J. E. G\'{o}mez-Correa},
journal = {Opt. Express},
number = {19},
pages = {33896--33910},
publisher = {Optica Publishing Group},
title = {Geometrical-light-propagation in non-normalized symmetric gradient-index media},
volume = {30},
month = {Sep},
year = {2022},
url = {https://opg.optica.org/oe/abstract.cfm?URI=oe-30-19-33896},
doi = {10.1364/OE.465957},
}

@article{Gomez-Correa:21,
author = {J. E. G\'{o}mez-Correa and A. L. Padilla-Ortiz and A. Jaimes-N\'{a}jera and J. P. Trevino and S. Ch\'{a}vez-Cerda},
journal = {Opt. Express},
number = {21},
pages = {33009--33026},
publisher = {Optica Publishing Group},
title = {Generalization of ray tracing in symmetric gradient-index media by Fermat's ray invariants},
volume = {29},
month = {Oct},
year = {2021},
url = {https://opg.optica.org/oe/abstract.cfm?URI=oe-29-21-33009},
doi = {10.1364/OE.440410},
}

@article{GomezCorrea2023,
  author = {Gómez-Correa, J. E. and Padilla-Ortiz, A. L. and Trevino, J. P. and Jaimes-Nájera, A. and Lozano-Crisóstomo, N. and Cornejo-Rodriguez, A. and Chavez-Cerda, S.},
  title = {Symmetric gradient-index media reconstruction},
  journal = {Optics Express},
  volume = {31},
  number = {18},
  pages = {29196--29212},
  year = {2023}
}

@article{Grzesik,
author = {   J. A.   Grzesik },
title = {Focusing Properties of a Three-parameter Class of Oblate, {Luneburg}-like Inhomogeneous Lenses},
journal = {Journal of Electromagnetic Waves and Applications},
volume = {19},
number = {8},
pages = {1005-1019},
year  = {2005},
publisher = {Taylor & Francis},
doi = {10.1163/156939305775526089},
URL = {https://doi.org/10.1163/156939305775526089},
}

@article{Navarro,
    author = {Mazzaferri, Javier and Navarro, Rafael},
    title = {Wide two-dimensional field laser ray-tracing aberrometer},
    journal = {Journal of Vision},
    volume = {12},
    number = {2},
    pages = {2-2},
    year = {2012},
    month = {02},
    issn = {1534-7362},
    doi = {10.1167/12.2.2},
    url = {https://doi.org/10.1167/12.2.2},
}

@article{Marchand:70,
author = {Erich W. Marchand},
journal = {J. Opt. Soc. Am.},
number = {1},
pages = {1--7},
publisher = {OSA},
title = {Ray Tracing in Gradient-Index Media},
volume = {60},
month = {Jan},
year = {1970},
url = {http://www.osapublishing.org/abstract.cfm?URI=josa-60-1-1},
doi = {10.1364/JOSA.60.000001},
}

@article{Marchand:88,
author = {Erich W. Marchand},
journal = {Appl. Opt.},
number = {3},
pages = {465--467},
publisher = {OSA},
title = {Rapid ray tracing in radial gradients},
volume = {27},
month = {Feb},
year = {1988},
url = {http://ao.osa.org/abstract.cfm?URI=ao-27-3-465},
doi = {10.1364/AO.27.000465},
}

@article{Marchand:72,
author = {E. W. Marchand},
journal = {Appl. Opt.},
number = {5},
pages = {1104--1106},
publisher = {OSA},
title = {Ray Tracing in Cylindrical Gradient-Index Media},
volume = {11},
month = {May},
year = {1972},
url = {http://ao.osa.org/abstract.cfm?URI=ao-11-5-1104},
doi = {10.1364/AO.11.001104},
}

@article{Moore:75,
author = {Duncan T. Moore},
journal = {J. Opt. Soc. Am.},
number = {4},
pages = {451--455},
publisher = {OSA},
title = {Ray tracing in gradient-index media},
volume = {65},
month = {Apr},
year = {1975},
url = {http://www.osapublishing.org/abstract.cfm?URI=josa-65-4-451},
doi = {10.1364/JOSA.65.000451},
}

@article{Streifer:71,
author = {W. Streifer and K. B. Paxton},
journal = {Appl. Opt.},
number = {4},
pages = {769--775},
publisher = {OSA},
title = {Analytic Solution of Ray Equations in Cylindrically Inhomogeneous Guiding Media. 1: Meridional Rays},
volume = {10},
month = {Apr},
year = {1971},
url = {http://ao.osa.org/abstract.cfm?URI=ao-10-4-769},
doi = {10.1364/AO.10.000769},
}

@article{Paxton:71,
author = {K. B. Paxton and W. Streifer},
journal = {Appl. Opt.},
number = {5},
pages = {1164--1171},
publisher = {OSA},
title = {Analytic Solution of Ray Equations in Cylindrically Inhomogeneous Guiding Media. Part 2: Skew Rays},
volume = {10},
month = {May},
year = {1971},
url = {http://ao.osa.org/abstract.cfm?URI=ao-10-5-1164},
doi = {10.1364/AO.10.001164},
}

@article{Buchdahl:73,
author = {H. A. Buchdahl},
journal = {J. Opt. Soc. Am.},
number = {1},
pages = {46--49},
publisher = {OSA},
title = {Rays in gradient-index media: separable systems},
volume = {63},
month = {Jan},
year = {1973},
url = {http://www.osapublishing.org/abstract.cfm?URI=josa-63-1-46},
doi = {10.1364/JOSA.63.000046},
}

@article{Sharma:82,
author = {Anurag Sharma and D. Vizia Kumar and A. K. Ghatak},
journal = {Appl. Opt.},
number = {6},
pages = {984--987},
publisher = {OSA},
title = {Tracing rays through graded-index media: a new method},
volume = {21},
month = {Mar},
year = {1982},
url = {http://ao.osa.org/abstract.cfm?URI=ao-21-6-984},
doi = {10.1364/AO.21.000984},
}

@article{Sharma:85,
author = {Anurag Sharma},
journal = {Appl. Opt.},
number = {24},
pages = {4367--4370},
publisher = {OSA},
title = {Computing optical path length in gradient-index media: a fast and accurate method},
volume = {24},
month = {Dec},
year = {1985},
url = {http://ao.osa.org/abstract.cfm?URI=ao-24-24-4367},
doi = {10.1364/AO.24.004367},
}

@article{Sakamoto:87,
author = {Toyokazu Sakamoto},
journal = {Appl. Opt.},
number = {15},
pages = {2943--2946},
publisher = {OSA},
title = {Ray trace algorithms for {GRIN} media},
volume = {26},
month = {Aug},
year = {1987},
url = {http://ao.osa.org/abstract.cfm?URI=ao-26-15-2943},
doi = {10.1364/AO.26.002943},
}

@article{Liu:23,
author = {Wanguo Liu and Yinshuan Ren and Fenghua Liu},
journal = {J. Opt. Soc. Am. A},
number = {1},
pages = {78--84},
publisher = {Optica Publishing Group},
title = {Analytical ray tracing based on Hamilton principal function and conjugate variable pairs},
volume = {40},
month = {Jan},
year = {2023},
url = {https://opg.optica.org/josaa/abstract.cfm?URI=josaa-40-1-78},
doi = {10.1364/JOSAA.477881},
}

@article{Chavez-Islas:25,
author = {V. Ch\'{a}vez-Islas and E. Esp\'{i}ndola-Ramos and F. H. Maldonado-Villamizar and S. Ch\'{a}vez-Cerda and J. E. G\'{o}mez-Correa},
journal = {Opt. Express},
number = {26},
pages = {54867--54878},
publisher = {Optica Publishing Group},
title = {Ray tracing method beyond idealized gradient-index lenses},
volume = {33},
month = {Dec},
year = {2025},
url = {https://opg.optica.org/oe/abstract.cfm?URI=oe-33-26-54867},
doi = {10.1364/OE.577925},
}

@article{Chavez2026generalized,
  title={Generalized vectorial ray tracing for optical analysis and design in freeform gradient-index media},
  author={Ch{\'a}vez-Islas, V and Esp{\'\i}ndola-Ramos, E and Maldonado-Villamizar, FH and Ch{\'a}vez-Cerda, S and G{\'o}mez-Correa, JE},
  journal={arXiv preprint arXiv:2606.22655},
  year={2026}
}

\end{document}